\documentclass[aps,twocolumn]{revtex4}
\usepackage{epsfig}
\usepackage{psfrag}
\newcommand{\be}{\begin{equation}}
\newcommand{\ee}{\end{equation}}
\newcommand{\ba}{\begin{array}}
\newcommand{\ea}{\end{array}}

\begin{document}

\title{2-Dimensional Polymers Confined in a Strip}

\author{Hsiao-Ping Hsu and Peter Grassberger}

\affiliation{John-von-Neumann Institute for Computing, Forschungszentrum J\"ulich,
D-52425 J\"ulich, Germany}

\date{\today}

\begin{abstract}
Single two dimensional polymers confined to a strip are studied by Monte
Carlo simulations. They are described by $N$-step self-avoiding random walks 
on a square lattice between two parallel hard walls with distance 
$1 \ll D \ll N^\nu$ ($\nu = 3/4$ is the Flory exponent). For the simulations 
we employ the pruned-enriched-Rosenbluth method (PERM) with Markovian 
anticipation. We measure the densities of monomers and of end points as 
functions of the distance from the walls, the longitudinal extent of the 
chain, and the forces exerted on the walls. Their scaling with $D$ and 
the universal ratio between force and monomer density at the wall are 
compared to theoretical predictions. 
\end{abstract}

\maketitle
\section{Introduction}

The behaviour of flexible polymers in a good solvent confined to different
geometries and in the presence of walls or other obstacles
have been studied for many years~\cite{deGen,Eisenriegler93}. A 
particularly simple geometry is the space between two parallel walls. 
For simplicity we shall only discuss here the case of walls without 
energetic effects, i.e. the walls play a purely geometric role.

An important theoretical prediction is that near such a wall the monomer 
density profile increases as $z^{1/\nu}$, where $z$ is the distance from the
wall ($z\ll D$, and $D$ is the width between the two parallel walls)  
and $\nu$ is the Flory exponent~\cite{deGen}. This should hold in any
dimension of space $d$. On the other hand it is intuitively obvious that 
the force exerted by the polymer onto the wall is proportional to the 
monomer density near the wall. The ratio between the two can be expressed
in terms of a universal amplitude ratio which has been calculated by 
Eisenriegler~\cite{Eisenriegler2} (using {\it conformal invariance} results of
Cardy et al.~\cite{cardy}) in $d=2$, 
and in any 
$d\le 4$ by Eisenriegler~\cite{Eisenriegler} by means of an $\epsilon$-expansion. 
Attempts to verify these detailed predictions by Monte Carlo simulations
in three dimensions~\cite{Webman,Ishinabe, Milchev,Joannis} have so far 
been without very convincing results. 
As far as we know, no attempt was made yet to verify them in $d=2$, and that
is where the present paper sets in.

\begin{figure}
  \begin{center}
\psfig{file=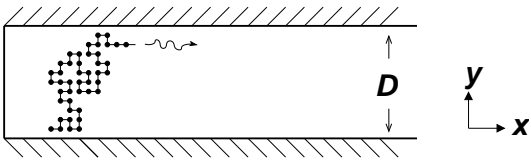,width=7.0cm, angle=0}
   \caption{Schematic drawing of a polymer chain growing inside a strip, 
            and with an additional wall added at $x=0$.
            Monomers are only allowed at lattice sites $x>0$ and 
            $1\leq y \leq D$.}
    \label{strip}
  \end{center}
\end{figure}

We study single polymer chains confined to a 2-d strip.
They are described by self-avoiding random walks (SAWs) of $N$ steps
on a square lattice between two hard walls with distance $D$ as shown
in Fig.~\ref{strip}. More precisely, monomers are supposed to sit on 
lattice sites and $D$ is the number of rows accessible to monomers, 
i.e. the walls are placed at $y=0$ and at $y=D+1$, and the monomers can 
be at $y=1,\ldots D$. We only consider the case where the Flory radius 
of a free chain of length $N$, $R_F\approx N^\nu$, is much larger than 
$D$. When using a chain growth algorithm, the polymer has then to grow
, after a short initial phase of $\sim D^{1/\nu}$ steps,
in either the positive or negative $x$-direction without possibility 
to change its orientation. This allows
us to use an additional wall at $x < x_0$ which forces all chains to 
grow into the positive $x$-direction. For $N\gg D^{1/\nu}$ this will
essentially reduce the partition sum by a constant factor, without
affecting any of the scaling laws or any of the detailed comparisons
with theoretical predictions. On the other hand, it simplifies the 
subsequent discussion.

The force exerted onto the wall is most straightforwardly expressed 
in terms of the work done when moving one of the walls, i.e. by the 
dependence of the free energy -- and thus also of the partition sum -- 
on $D$,
\be
   F = k_BT\; {\partial\ln Z_N \over \partial D}\;,        \label{F}
\ee
where we have introduced a dummy temperature $T$ which can take any 
positive value. The partition sum $Z_N$ is just the number of $N$-step 
SAWs in the strip starting from a given $x$, but summed over all values
$1\leq y_0 \leq D$ of the $y$-component of the starting point.

The partition sum of a free SAW in infinite volume scales for $N\to\infty$ 
as $Z_N = \mu_\infty^{-N} N^{\gamma-1}$ with $\mu_\infty$ being the 
critical fugacity per monomer, and with $\gamma=43/32$ being a universal 
exponent. In contrast, the partition function on a strip scales as
\be
       Z_N \sim \mu_D^{-N} \; ,          \label{ZN}
\ee
without the power correction and with $\mu_D$ scaling for large $D$
as 
\be
      \mu_D-\mu_\infty \approx a D^{-1/\nu} \; .  \label{MUD}
\ee
with $a$ being a non-universal amplitude.
The force per monomer is then obtained as 
\be
   f = F/N \sim k_BT \; D^{-1-1/\nu}\;.        \label{ff}
\ee

Standard fixed-length Markov Chain Monte Carlo simulations do not 
give estimates of the partition sum or of the free energy, so that 
Eqs.~(\ref{F}-\ref{ff}) cannot be used directly. This has led to 
algorithms specifically designed for estimation of forces \cite{dickman},
but employing the pruned-enriched-Rosenbluth method (PERM)~\cite{g97}
one can use Eqs.~(\ref{F}-\ref{ff}) directly.

Using PERM with $k$-step Markovian anticipation~\cite{Frauenkron98,
Frauenkron99,Caracciolo}, we measured the partition
sum directly and estimated the dependence of the monomer fugacity on the 
width $D$. In the same simulations also the monomer density profile, the 
end-to-end distance along the strip, and the density profile of chain ends
are measured.

Details of the simulation method are given in the next section, results and 
their comparison with theoretical predictions are discussed in Sec.~III.

\section{Algorithm: PERM with $k$-step Markovian anticipation}

PERM \cite{g97} is a chain growth algorithm with population control. 
Polymer chains are built like random walks by adding one monomer at each 
step. We use a Rosenbluth like bias for self-avoidance. As usual, this 
bias is compensated by a weight factor \cite{rosenbluth}, i.e.
each sample configuration should be given a weight. But actually
we use a stronger bias which in addition suppresses dense configurations 
and samples more finely relatively open chain configurations, called 
{\it Markovian anticipation} in \cite{Frauenkron99}.

In $k$-step Markovian anticipation, we let the additional bias in the 
next step depend on the last $k$ steps made before. Let us denote the 
$2d$ directions on a $d$-dimensional hypercubic lattice by $s=0, \ldots, 2d-1$.
All possible $(k+1)$-step configurations ($k$ previous steps $i=-1,-2,\ldots -k$
plus one future step $i=0$) are then indexed by 
\be
   {\bf S} = (s_{-k},\ldots,s_0)=({\bf s},s_0) \; .
\ee
In addition, choose an integer $m\approx 100$ (the precise value is not 
important).
Either during an auxiliary run or during the early stages of the present 
run, we obtained a histogram $H_m$ such that $H_m({\bf S})$ is the accumulated
weight at chain length $n+m$ of those chains which had configuration ${\bf S}$ during 
steps $n-k,n-k+1,\ldots n$. Since we sample uniformly, $H_0({\bf S})$ is 
independent of ${\bf S}$. Thus $H_m({\bf S})$ with $m>0$ indicates how ``successful"
is a configuration ${\bf S}$ after $m$ more steps. In importance sampling we want 
each chosen direction to have in average the same later success. Therefore
we choose the next step with probability
\be
      p(s_0|{\bf s})=\frac{H_m({\bf s},s_0)}
             {\sum_{s_0'=0}^{2d-1} H_m({\bf s},s_0')} \;,
   \label{markova}
\ee
In our simulations we choose $k=9$ and $m=100$. We accumulate contributions to
$H$ only for $n+m>300$, and we apply Eq.~(\ref{markova}) only for chain lengths $> k$
(for chain lengths $<k$ there is not yet enough history to condition upon).

Accumulating the histogram only for $n\gg 1$ is suggested by the fact that only
for large $n$ the anisotropic bias is fully developed \cite{Frauenkron99}.
This anisotropy makes the histogram strongly dependent on $D$. We found 
that using for all $D$ only the histogram obtained for free chains, i.e. for
$D=\infty$, gives nearly the same efficiency. This is quite different from the 
case where the anisotropy is not due to geometry, but is due to stretching of 
the polymer. In the latter case, simulations with markovian anticipation 
become much more efficient with increased stretching \cite{unpub}. This is not 
the case for the present problem where the anisotropy is due to geometric 
constraints, for which PERM with Markovian anticipation is however still the most 
efficient known simulation method by far.

\section{Results}

Before presenting our results, let us stress that we have several possibilities
for checking our algorithm. For very large $D$ we can compare our estimates of $\mu$
with the very precise estimate $\mu_\infty=0.37905228$ \cite{Guttmann88,Guttmann96}.
For $D\le 11$ we can compare with exact transfer matrix calculations of 
\cite{Burkhardt}. And for $D\le 2$ we can even solve the problem analytically.

\begin{figure}
  \begin{center}
\psfig{file=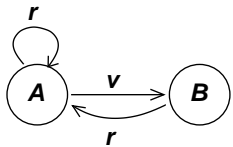,width=3.0cm, angle=0}
   \caption{Graph accepting the regular grammar of vertical (`v') and right (`r')
    moves for $D=2$. The node labelled `A' is the start node~\cite{hopcroft}.
    If one wants to distinguish also between `u' and `d' moves, the graph is
    somewhat more complicated and contains also a transient part. This is skipped
    here for simplicity.}
    \label{graph}
  \end{center}
\end{figure}

For $D=1$, the polymer can only grow in a straight configuration, giving $\mu_{(D=1)}=1$.
For $D=2$, each step can be either up (u), down (d), or to the right (r). After an `u' 
or `d' move, the next step has to be `r', while any move is possible after `r'. Lumping
moves `u' and `d' together into a vertical move (`v'), we see that the set of all 
possible configurations forms a regular language \cite{hopcroft} with the associated 
graph shown in Fig.~\ref{graph}. The partition sum for chains of length $N$ is just 
twice the $(N+1)$-st Fibonacci number,
\be
       Z_N(D=2)=2F_{N+1}  \quad (N\ge 1)\; ,   \label{ZD3}
\ee
where $F_0=F_1=1$ and $F_N=F_{N-1}+F_{N-2}$. Thus the critical monomer fugacity for
$D=2$ is the inverse of the golden mean,
\be
     \mu_{D=2} = g^{-1} \equiv \frac{\sqrt{5}-1}{2} = 0.61803\ldots.     \label{Golden}
\ee
In this case, we can also show that markovian anticipation gives the optimal bias.
Markovian anticipation corresponds in this case to $p_r:p_v = g:1$ if a vertical
move is allowed, and $p_r:p_v = 1:0$ else. This choice leads to weights which 
oscillate between two values, thus no population control (pruning/cloning) is 
needed \cite{footnote}. All this is verified in our simulations, which serves 
thus as a test for our algorithms.

\begin{figure}
  \begin{center}
\psfig{file=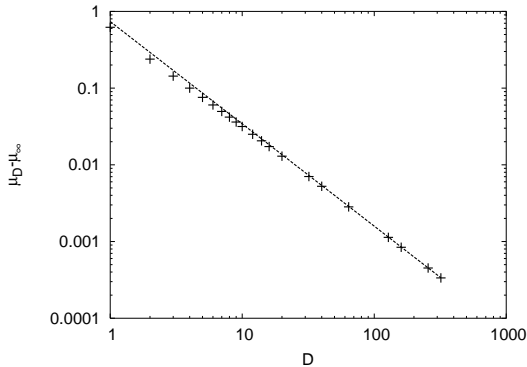,width=5.0cm, angle=270}
   \caption{Log-log plot of $\mu_D-\mu_\infty$ against $D$. The dashed line
            is $\mu_D-\mu_\infty = 0.737 D^{-1/\nu}$ with $\nu=3/4$,
            as predicted by Eq.~(\ref{MUD})}
    \label{mu}
  \end{center}
\end{figure}

For $D>2$, we used simulations.
We simulated strip widths up to $D=320$ and chain length between $N=3000$
(for $D=2$) and $N=125,000$ (for $D=320$). Critical fugacities are 
determined by plotting $Z_N \mu_D^N$ against $\log N$ and demanding that
these curves become horizontal for large $N$. Results are shown in Fig.~\ref{mu},
where we plot $\mu_D-\mu_\infty$ with $\mu_\infty=0.37905228$~\cite{Guttmann96}. 
They are in perfect agreement with the theoretical prediction of Eq.~(\ref{MUD}), 
and provide the estimate $a=0.7365\pm 0.0007$. In addition, $\mu_D$ can be compared 
for $D=3$ to $12$ with the transfer matrix results of \cite{Burkhardt}. For all 
$D$, the values agree for at least six digits.
 
\begin{figure}
  \begin{center}
\mbox{(a)}
\psfig{file=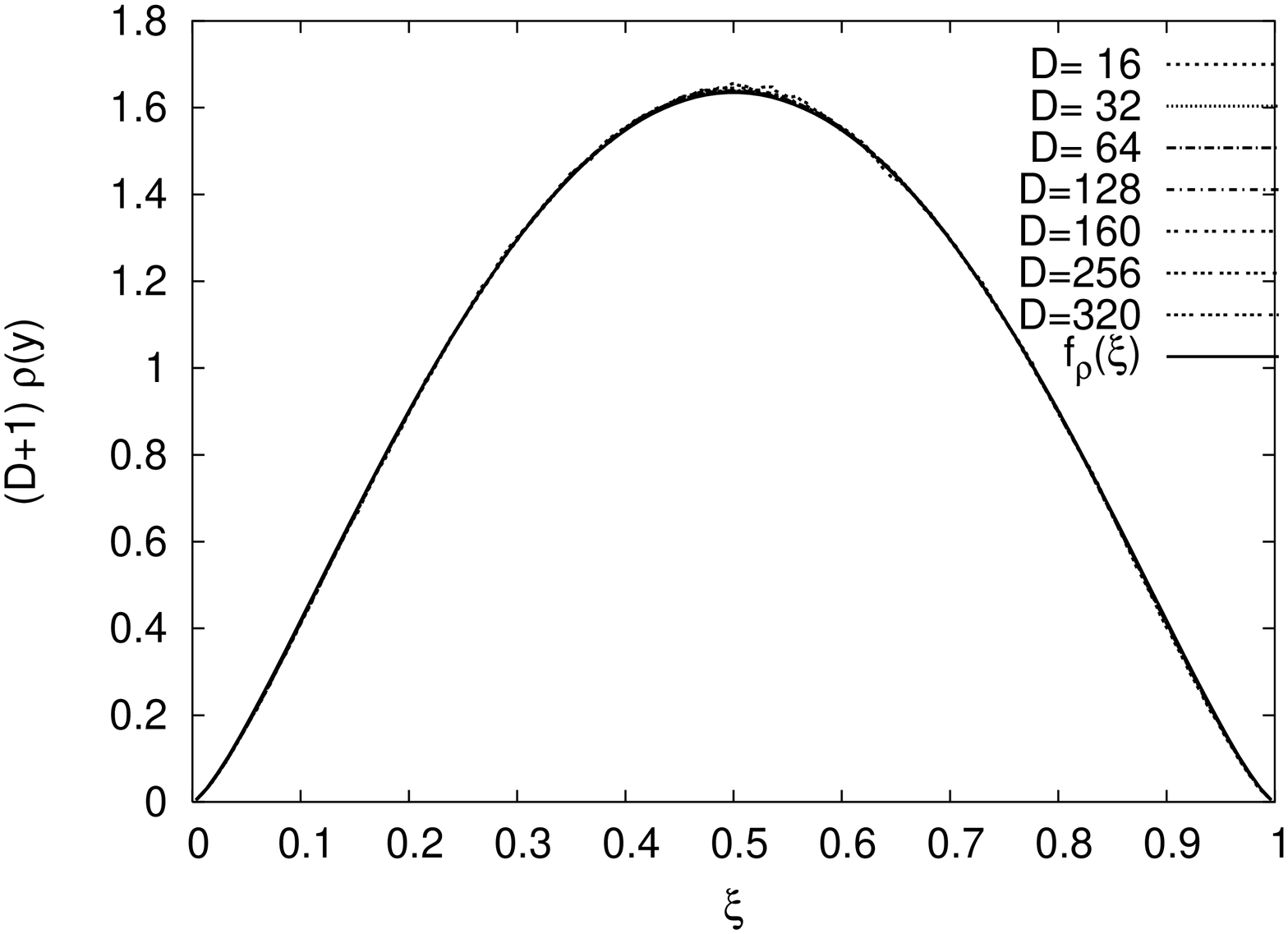,width=5.0cm, angle=270} \\
\mbox{(b)}
\psfig{file=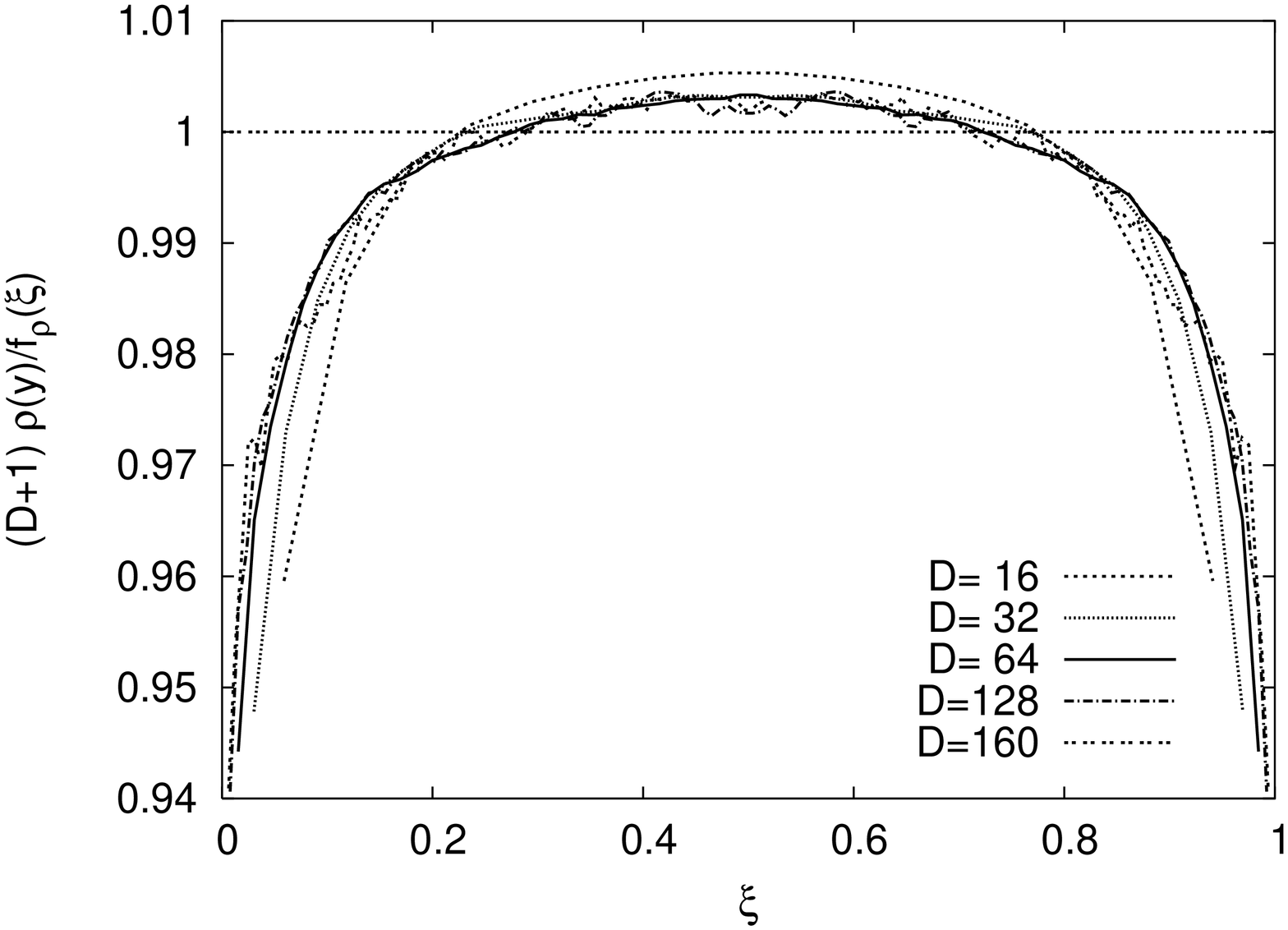,width=5.0cm, angle=270} \\
\mbox{(c)}
\psfig{file=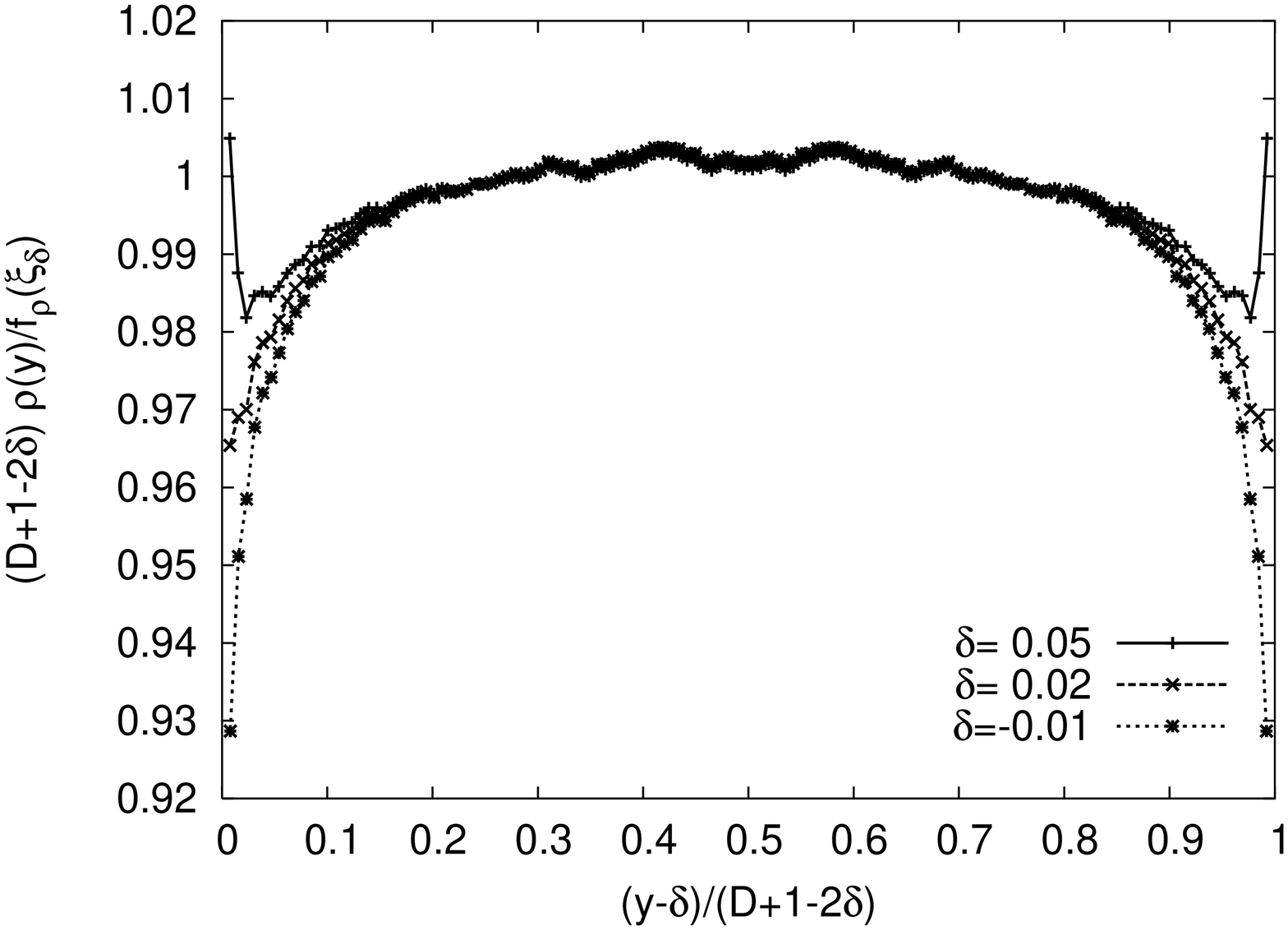,width=5.0cm, angle=270}
   \caption{(a) Rescaled values of the monomer density, $(D+1)\; \rho(y)$
       against $\xi=y/(D+1)$. Also plotted is the function $f_\rho(\xi) = 
       10.38\;(\xi(1-\xi))^{4/3}$.\\
     (b) The same values (for $D\le 160$), but divided by $f_\rho(\xi)$.
       In this panel we do not display our data for the largest lattices,
       since they are too noisy and would just blur the picture. They do 
       however show the same trend as the data for $D\le 160$. \\
     (c) The data for $D=128$ plotted against a modified scaling variable,
       $\xi_\delta = (y-\delta)/(D+1-2\delta)$, divided by 
       $f_\rho(\xi_\delta)$, for three different values of $\delta$. }
    \label{rhoy}
  \end{center}
\end{figure}

\begin{figure}
  \begin{center}
\psfig{file=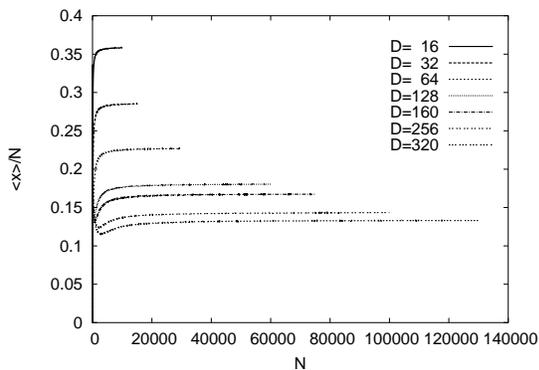,width=5.0cm, angle=270}
   \caption{End-to-end distance divided by $N$, $\langle x\rangle/N$, plotted against $N$ for
     various values of $D$. }
    \label{x}
  \end{center}
\end{figure}

\begin{figure}
  \begin{center}
\psfig{file=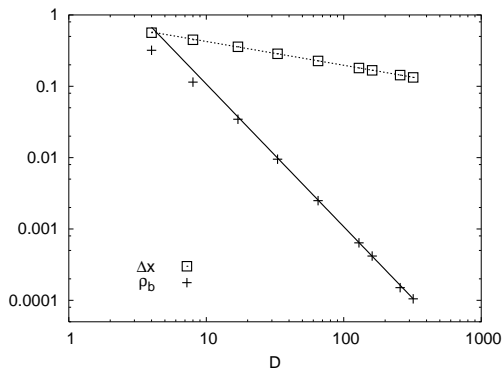,width=5.0cm, angle=270}
   \caption{log-log plots of $\Delta x$ and of the boundary density $\rho_b$ versus $D$.
   The dashed line is $0.915\;D^{-1/3}$ and the solid line is $10.75\; D^{-2}$. }
    \label{scal}
  \end{center}
\end{figure}

\begin{figure}
  \begin{center}
\mbox{(a)}
\psfig{file=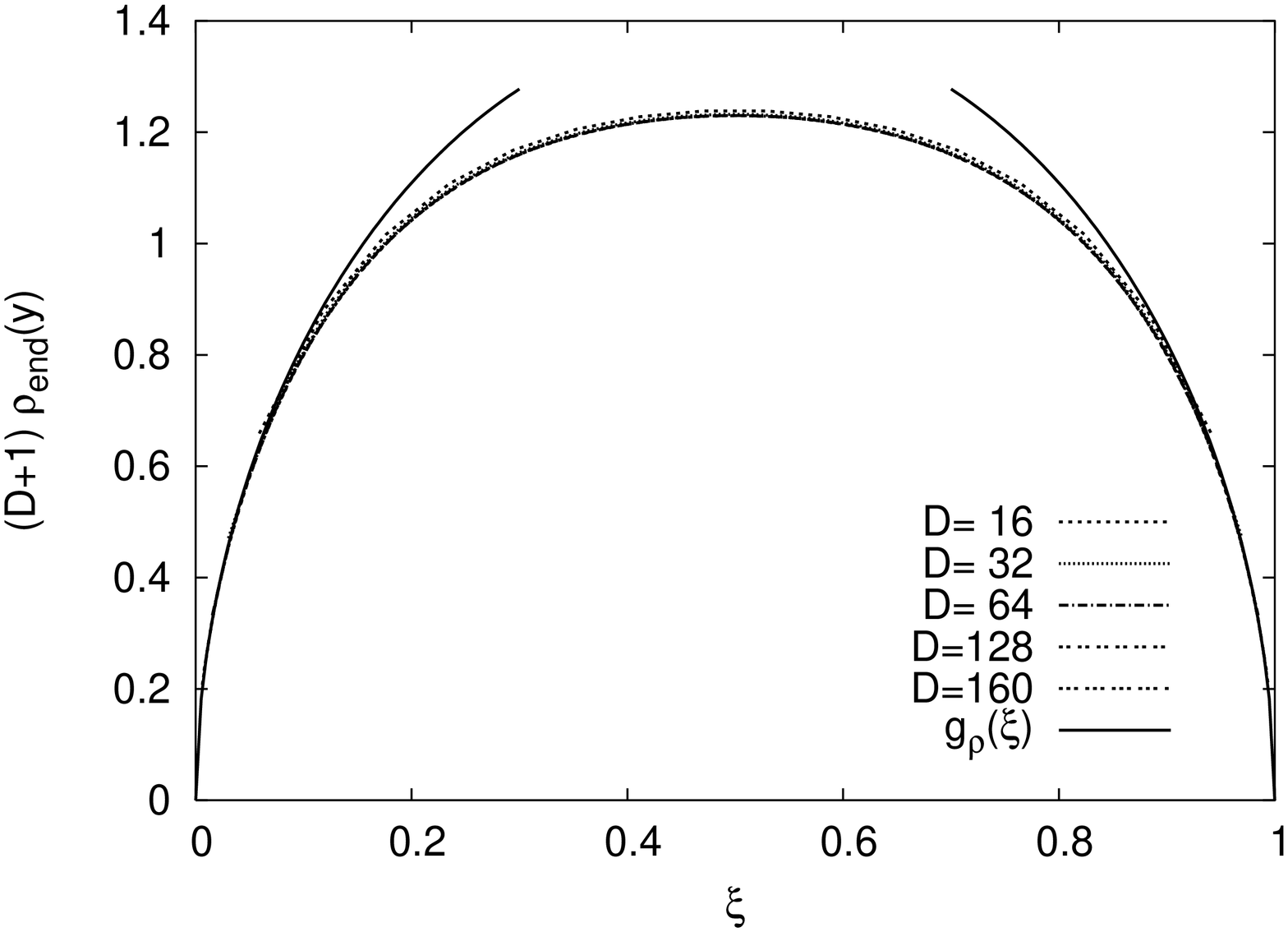,width=5.0cm, angle=270} \\
   \caption{Rescaled values of the probability $\rho_{end}(y)$ that the chain
       end is at the distance $y$ from a wall against $\xi=y/(D+1)$. The topmost
       (interrupted) line is the function $g_\rho(\xi)=2.85\;(\xi(1-\xi))^{25/48}$.}
            \label{rhoend}
  \end{center}
\end{figure}

As we said in the introduction, we expect the force $f$ onto the wall to be 
proportional to the monomer density $\rho(y)$ near the wall. The precise relation 
is given by Eisenriegler~\cite{Eisenriegler} 
\be
     \lim_{y \rightarrow 0} k \frac{\rho(y)}{y^{1/\nu}}=
     B \frac{f}{k_BT} \; .       \label{rhof}
\ee
Here the non-universal amplitude $k$ relates the end-to-end distance of a 
free SAW to the chain length, $k=R_x^{1/\nu}/N= 0.5297\pm 0.0002$ 
\cite{LiMadrasSokal} for the
square lattice. On the other hand, $B$ is a universal number. For ideal
chains $B=2$, while for chains with excluded volume in $4-\epsilon$ 
dimensions one has $B \approx 2(1-b_1 \epsilon)$ with $b_1=0.075$
\cite{Eisenriegler}. The latter is of course of dubious value for $d=2$
(where it would give $B\approx 1.7$),
but conformal invariance leads to the supposedly exact value $B=2.01$ 
in $d=2$ \cite{cardy,Eisenriegler2}. 

The monomer density is normalized such that $\sum_{y=1}^D \rho(y)=1$. 
According to Eq.~(\ref{rhof}) it should scale as $y^{1/\nu}$ near the 
walls (this holds in any dimension) with $\nu=3/4$. Surprisingly, we found 
that the simplest ansatz generalizing this power law to all $y\in[1,D]$, 
\be
 \rho(y)= \frac{1}{D+1}f_\rho(\xi)\equiv \frac{1}{D+1}A(\xi(1-\xi))^{4/3} \; , 
\; \xi={y\over D+1} \; , \label{frho}
\ee
with $A=\Gamma(14/3)/\Gamma^2(7/3)=10.38$ gives already an excellent 
fit (Fig.~\ref{rhoy}a) in the limit $D \rightarrow \infty$.
There are small deviations invisible in Fig.~\ref{rhoy}a but clearly 
seen when plotting $\rho(y)/f_\rho(\xi)$ (see Fig.~\ref{rhoy}b), but 
they seem to vanish slowly in the limit $D\to\infty$ (see Fig.~\ref{rhoy}b).

Assuming the latter, i.e. assuming that Eq.~(\ref{frho}) becomes exact 
asymptotically, the universal number $B$ could be estimated by using 
Eqs.~(\ref{F}), (\ref{ZN}), (\ref{MUD}), and (\ref{rhof}) which together 
would give
\be
   A = \lim_{D\to\infty}\lim_{y \to 0} \frac{D^{7/3}\rho(y)}{y^{4/3}}
   =\frac{4}{3}\frac{Ba}{k\mu_\infty} \; .
\ee
Inserting the above numbers gives $B= 2.122\pm 0.002$ which is definitely 
larger than the value predicted in \cite{cardy,Eisenriegler2}, by some
50 standard deviations. 

In an alternative scenario we could assume that the deviations from 
Eq.~(\ref{frho}) seen in Fig.~\ref{rhoy}b do not vanish in the limit
$D\to\infty$. In that case one should also allow for a modified scaling 
variable 
\be
   \xi_\delta=(y-\delta)/(D+1-2\delta)
\ee 
with an unknown small (non-universal) parameter $\delta$. Values of 
$(D+1-2\delta) \rho(y)/f_\rho(\xi_\delta)$ for $D=128$ and three 
different values
of $\delta$ are plotted in Fig.~\ref{rhoy}c. Similar results are obtained 
for other $D$. Combining them, we see that best scaling (i.e. least 
dependence of $D$ for small $y$) is obtained for $\delta\approx 0.02$.
For this value of $\delta$ we have 
\be
   \lim_{y\to 0,\;D\to\infty} D^{7/3}\rho(y)/y^{4/3} = (0.95\pm 0.02)\times A\;.
       \label{A_delta}
\ee
The large uncertainty in this estimate reflects the rather steep slope
of the central curve in Fig.~\ref{rhoy}c as $\xi_\delta\to 0$ and the 
related uncertainty in the best estimate of $\delta$~\cite{footnote2}. If we 
accept Eq.~(\ref{A_delta}), we obtain $B=2.04\pm 0.04$, which is in perfect 
agreement with \cite{cardy,Eisenriegler2}. Thus we have two scenarios.
Both imply very large corrections to scaling. While the a priori simpler 
scenario would be in conflict with the theoretical prediction, this prediction
suggests that indeed the second scenario is correct, which is our
preferred solution. 

  In Fig.~\ref{x}, we plot the end-to-end distance per monomer 
$\langle x\rangle/N$ versus $N$ for various widths $D$. These curves become
horizontal as $N \to \infty$, i.e. $\langle x\rangle$ increases indeed
linearly with $N$, $\lim_{N\to\infty} \langle x\rangle/N = \Delta x$. In 
order to find how $\Delta x$ scales with $D$, we plot it in Fig.~\ref{scal}
on a doubly logarithmic scale. As indicated by the dashed line, it is 
fitted perfectly by the theoretical prediction \cite{Eisenriegler}
\be
       \langle x\rangle/N \sim D^{1-1/\nu} = D^{-1/3} \;.   \label{xn}
\ee 

We also estimated the density of wall contacts $\rho_b$ (number of monomers 
at $y=1$ or at $y=D$, divided by $2\langle x\rangle$). For each fixed value 
of $D$, this becomes independent of $N$ as $N\to\infty$. The asymptotic
values, obtained from plots similar to Fig.~\ref{x}, are also plotted in 
Fig.~\ref{scal}. The full line corresponds to the very simple
prediction \cite{Eisenriegler}
\be
      \rho_b \sim D^{-2}    \label{rhob}
\ee
which is independent of $\nu$ and indeed holds also for Gaussian chains.
Eq.~(\ref{rhob}) can be easily understood, in terms of the pressure
exerted onto the wall: 
\be
   \rho_b \sim p = {Nf\over \langle x\rangle} \sim D^{-1-1/\nu+1/3} = D^{-2} \; .
\ee

Finally, we show in Fig.~\ref{rhoend} the distribution $\rho_{end}(y)$ of chain 
ends. We see that $\rho_{end}(y)$ scales for large $D$ (and for $N\gg D^{1/\nu}$,
of course). Theoretically it is predicted that \cite{Cardy-Redner}
\be
   \rho_{end}(y) \sim y^{25/48}
\ee
for $y \ll D$. We see that this is indeed verified (the full lines in
Fig.~\ref{rhoend} correspond to the prediction). But in contrast to the
monomer density which was described fairly well for all $y$ by the product of
the power laws holding near the two walls, the same definitely does not hold for
$\rho_{end}(y)$. There the function $g_\rho(\xi) = const\; (\xi(1-\xi))^{25/48}$
does not describe the behaviour away from the walls.

\section{Summary}

We have shown that we could simulate 2-d polymers, modelled as self-avoiding 
walks, with chain length up to 125000 on strips of widths up to 320. This 
was possible using the PERM algorithm with Markovian anticipation. The fact
that PERM gives by default very precise estimates of free energies allowed
us to measure precisely the forces exerted onto the walls, by measuring how 
the critical fugacities depend on the width of the strips. We verified all
critical scaling laws predicted for this problem, including the scaling of 
monomer and end point densities near the walls and the scaling of the total
pressure with chain length and with strip width.

The only prediction for which we found possibly disagreement is for the 
universal amplitude ratio $B$ defined in Eq.~(\ref{rhof}). A scenario based
on some minimal assumption about scaling functions and corrections to 
scaling gives an estimate higher than the prediction by
some fifty standard deviations. But a different scenario, maybe less 
plausible a priori but not very unlikely either, gives perfect agreement
with the prediction. This illustrates again that one should be very 
careful about corrections to scaling, and that even very precise simulations 
do not always give unique answers when their analysis is not guided 
by a reliable theory.

Previous simulations of 3-d polymers between two parallel planar walls had
indicated that also there the value of $B$ might be larger than predicted, 
but those simulations had very large uncertainties. Using PERM we can simulate 
fairly easily much longer chains with rather high statistics. Results of such 
3-d simulations will be given elsewhere.

Acknowledgements: We thank Profs. Erich Eisenriegler and Ted Burkhardt for 
valuable discussions, and Walter Nadler for carefully reading the manuscript.


\begin{thebibliography}{99}
\bibitem{footnote}
   An analogous result holds for any regular language with $m$ non-transient nodes 
   in its minimal graph (or equivalently for a Markov source with $m$ states): 
   Markovian anticipation leads to weights which take on $m$ different values,
   and thus no population control is needed to keep these weights within finite
   non-zero limits.
\bibitem{footnote2} Plotting $\rho(y)$ on a log-log plot would not help
either. The large scale seen in Fig.~4c implies that even the scaling
exponent $1/\nu$ would be estimated wrongly from such a plot, unless one
has much more complete data.
\end{thebibliography}
\end{document}